\documentclass[aps,pra,superscriptaddress,twocolumn]{revtex4}
\usepackage[utf8]{inputenc}
\usepackage{braket,textcomp}
\usepackage[usenames]{color}
\usepackage[dvipsnames]{xcolor}
\usepackage[breaklinks, colorlinks=true, urlcolor=blue, 
anchorcolor=blue, citecolor=blue, filecolor=blue, linkcolor=blue, 
menucolor=blue, linktocpage=true, pdfproducer=medialab, 
pdfa=true]{hyperref}
\usepackage{amssymb,amsmath,bbm,mathcomp}
\usepackage{amsfonts,amsthm,subfigure,nccmath}
\usepackage{braket}
\usepackage{times}
\usepackage{mathrsfs}
\usepackage[mathscr]{euscript}
\usepackage{calligra}
\usepackage{graphicx,times}
\usepackage{epsfig}
\usepackage{caption}
\usepackage{cancel}
\usepackage{pifont}
\usepackage{enumerate}
\usepackage{mhsetup}
\usepackage{mathtools}
\usepackage{bm}
\usepackage{empheq}
\newcommand{\defi}{\coloneqq}

\newcommand{\HH}{\mathcal H}
\newcommand{\Id}{\hat{\mathbb I}}
\newcommand{\scalar}[2]{\langle{#1}|{#2}\rangle}
\newcommand{\exval}[1]{\langle{#1}\rangle}

\newcommand{\clo}{\mbox{C}}
\newcommand{\smg}{{_{{\Gamma}\,}}}
\newcommand{\smc}{{_{{C}\,}}}
\newcommand{\proj}[1]{\ket{#1}\!\!\bra{#1}}
\newcommand{\dket}[1]{\ket{#1}\!\rangle}
\newcommand{\dbra}[1]{\langle\!\bra{#1}}

\graphicspath{{figure/}}

\begin{document}
\title{There is one time}
\author{Caterina Foti}
\address{Dipartimento di Fisica e Astronomia, Universit\`a di Firenze, I-50019,
Sesto Fiorentino (FI), Italy}
\address{INFN, Sezione di Firenze, I-50019, Sesto Fiorentino (FI), Italy}
\author{Alessandro Coppo}
\address{Dipartimento di Fisica e Astronomia, Universit\`a di Firenze, I-50019,
Sesto Fiorentino (FI), Italy}
\address{INFN, Sezione di Firenze, I-50019, Sesto Fiorentino (FI), Italy}
\author{Giulio Barni}
\address{Dipartimento di Fisica e Astronomia, Universit\`a di Firenze, I-50019,
Sesto Fiorentino (FI), Italy}
\author{Alessandro Cuccoli}
\address{Dipartimento di Fisica e Astronomia, Universit\`a di Firenze, I-50019,
Sesto Fiorentino (FI), Italy}
\address{INFN, Sezione di Firenze, I-50019, Sesto Fiorentino (FI), Italy}
\author{Paola Verrucchi}
\address{ISC-CNR, UOS Dipartimento di Fisica, Universit\`a di Firenze, I-50019,
Sesto Fiorentino (FI), Italy}
\address{Dipartimento di Fisica e Astronomia, Universit\`a di Firenze, I-50019,
Sesto Fiorentino (FI), Italy}
\address{INFN, Sezione di Firenze, I-50019, Sesto Fiorentino (FI), Italy}
\date{\today}
\title{There is only one time}

\begin{abstract} We draw a picture of physical systems that allows us to 
recognize what is this thing called "time" by requiring consistency not 
only with our notion of time but also with the 
way time enters the fundamental laws of Physics, independently of one 
using a classical or a quantum description. Elements of the picture are 
two non-interacting and yet entangled quantum systems, one of which 
acting as a clock, and the other one doomed to evolve. The setting is 
based on the so called "Page and Wootters (PaW) mechanism"
\cite{PageW83}, and updates~\cite{GiovannettiLM15,MarlettoV17,MacconeS20},
with tools from Lie-Group \cite{Gilmore12} 
and large-$N$ quantum approaches
\cite{Lieb73,Berezin78,Yaffe82,BrezinW93,CoppoCFVsoco20}.
The overall scheme is quantum, but the theoretical framework 
allows us to take the classical limit, either of the clock 
only, or of the clock and the evolving system altogether;
we thus derive the Schr\"odinger equation in the first case, and the 
Hamilton equations of motion in the second one.
Suggestions about possible links with general relativity and 
gravity are also put forward.
%This work shows that there is no such thing as a "quantum 
%time", possibly opposed to a "classical" 
%one: there is only one time, and it is a manifestation of entanglement. 

\end{abstract} 
\maketitle

\section{Introduction} 
\label{s.introduction} 

The notion of time is deeply rooted into our perception of reality, 
which is why, for centuries, time has entered Physics
as a fundamental ingredient that is not to be questioned. 
Then, general relativity (GR) and quantum mechanics (QM) 
intervened in opposite directions: GR gave time 
the same status of position, while QM made time a parameter, 
external to the theory and not recognizable as an observable.  While the 
introduction of "spacetime" in 
GR appears as an elegant intuition, fully consistent with classical 
physics, the fact that time cannot be treated as any other observable 
in QM is disturbing. As a 
consequence, discussions about the role of time in QM have been 
developed, leading to 
different proposals on how to overcome what seems a serious 
inconsistency of the theory. Reporting upon these discussions goes 
beyond the scope of this paper; therefore, in what follows we will only 
refer to the proposal that provides our starting point. 
This was introduced by D.~N.~Page and W.~K.~Wotters in 1983 
\cite{PageW83} to formalize the idea that the expression
"at a certain time $t$" should be understood as "conditioned to a
clock being in a state labeled by a certain value $t$". This 
proposal, to which we will refer as the "Page and Wootters (PaW) 
mechanism", is based upon three assumptions: {\it i)} the 
clock does not interact with the system to which it provides the 
parameter $t$, but {\it ii)} it is entangled with it; moreover, 
{\it 
iii)} clock and system together are in an eigenstate of the total 
Hamiltonian (with eigenvalue that can be set equal to zero, for the 
sake of simplicity and without loss of generality). The Paw mechanism 
has been extensively used, and its assumptions scrutinized, in the 
recent literature, both from the theoretical and the experimental viewpoint
\cite{GambiniPP04group,GambiniEtal09,MorevaEtal14,
GiovannettiLM15,MarlettoV17,LeonM17,MorevaEtal17,BryanM17,GourEtal18,MendesP19,
SmithA19,FavalliS20,MacconeS20,CastroRuizEtal20}.

Most discussions about time in QM are aimed at understanding what is the 
status of time in the quantum description, as if 
there were no problem as far as one stays classical.
However, if one believes that there do not exist quantum 
systems and classical ones, but rather that some quantum systems 
behave in a way that, under certain conditions, is efficiently described 
by the laws of classical physics, than there must be just one time.
In other terms, the procedure used to identify what time is 
in QM must have a well defined classical limit, fully 
consistent with classical physics and the way time 
enters the classical equation of motion.
In this work we construct such a procedure, and demonstrate 
that it consistently produces not only the Schr\"odinger equation for quantum 
systems, but also the Hamilton equations of motion (e.o.m.) for 
classical ones, with the parameter playing the role of time being the 
same in both cases. We tackle the quantum-to-classical crossover via 
the large-$N$ approach based on Generalized 
Coherent States (GCS) from Refs.~\cite{Lieb73,Berezin78,Yaffe82,BrezinW93,CoppoCFVsoco20}, 
where it is demonstrated that the theory describing a quantum 
system for which GCS can be constructed flows into a well defined 
classical theory if few specific conditions upon its GCS hold in the 
$N\to\infty$ limit ($N$ quantifies the number of microscopic quantum 
components, sometimes referred to as the number of degrees of freedom or 
dynamical variables, in the literature). By "classical 
limit" we will hereafter mean the large-$N$ limit with the above 
conditions on GCS enforced.

The paper is organized as follows.
In Secs.~\ref{s.non_interacting} and
\ref{s.letscallit-phi} we consider a completely quantum description.
In particular, in Sec.~\ref{s.non_interacting} we define the 
overall system as made of two non-interacting and entangled objects, 
dubbed {\it clock} and {\it evolving system} (or just {\it system} 
whenever possible), while in Sec.~\ref{s.letscallit-phi} we introduce a 
parametric representation for writing the entangled state of such system 
with GCS for the clock; this allows us to identify a real 
parameter $\varphi$ whose features make it a good candidate to 
represent time.
In Sec.~\ref{s.letscallit-t} we take the classical limit for the clock 
only, and derive an equation for the physical states of the 
system which is the Schr\"odinger equation, once 
the above mentioned parameter $\varphi$ is given the role of time.
In Sec.~\ref{s.emergence}, we take the classical limit 
of the evolving system too, and get to our most relevant result, 
namely that the Hamilton e.o.m. of classical physics are
consistently derived, with the same parameter $\varphi$ as
time, and the Planck's constant appearing in the classical Poisson 
brackets, as a footprint of their original quantum nature. 
Finally, in Sec.~\ref{s.conc-disc} we discuss our results and suggest 
some possible developments.

\section{Entangled, and yet non-interacting}
\label{s.non_interacting}

We consider a composite quantum system $\Psi=C+\Gamma$, with $C$ the clock
and $\Gamma$ the evolving system; 
we assume that $\Psi$ is isolated, with Hamiltonian $\hat H$, and
in a pure state $\dket{\Psi}$ which is entangled w.r.t. the partition 
$C$ and $\Gamma$; as in Ref.~\cite{GiovannettiLM15}, the double braket 
indicates states in
${\cal H}_\Psi={\cal H}_\smc\otimes{\cal H}_\smg$, 
with ${\cal H}_*$ the Hilbert space of $*=\Psi,C,\Gamma$.
Referring to the PaW mechanism,
we assume that 
\begin{equation}
\hat H\dket{\Psi}=0~,
\label{e.deWitt}
\end{equation}
and take $C$ and $\Gamma$ non-interacting, i.e.
\begin{equation}
\hat{H}=\hat{H}_\smc\otimes\Id_\smg-\Id_\smc\otimes\hat{H}_\smg~,
\label{e.H}
\end{equation}
where the irrelevant minus sign in front of the term acting on $\Gamma$ is our choice for the sake of a lighter notation.
In view of dealing with a parameter that must be continuous to 
represent time, we resort to a parametric representation (see 
supplementary material) of $\dket{\Psi}$ with GCS for the clock
~\cite{BarniBS15,CoppoMS19,Foti_Phd19}, and
write
\begin{equation}
\dket{\Psi}=\int_{{\cal M}_\smc} d\mu(\bm\Omega)
\chi(\bm\Omega)\ket{\bm\Omega}\otimes\ket{\phi(\bm\Omega)}~,
\label{e.Psi_para}
\end{equation}
where $\ket{\bm\Omega}$ are the GCS 
defined via the group-theoretical construction~\cite{Perelomov72,Gilmore72}
for the Lie group ${\cal G}_\smc$ associated with the algebra 
$\mathfrak{g}_\smc$ 
to which the Hamiltonian $\hat{H}_\smc$ belongs, and $\chi(\bm\Omega)$ can 
be chosen real without loss of generality. 
The $M$-tuples
$\bm\Omega=(\Omega_1,\Omega_2...\Omega_M)$, with
$\Omega_m\in\mathbb{C}~\forall m$, identify points on
$\cal{M}_\smc$, which is a $2M$-dimensional manifold
with a simplectic structure, and
$M$ related to the dimension of $\mathfrak{g}_\smc$. The measure 
$d\mu(\bm\Omega)$ is invariant w.r.t. the elements of ${\cal G}_\smc$ and 
ensures that GCS form a complete set upon ${\cal H}_\smc$, thus 
providing a resolution of the identity. The positive function 
$\chi^2(\bm\Omega)$ is a normalized probability distribution on ${\cal M}_\smc$, 
and $\ket{\phi(\bm\Omega)}\in{\cal H}_\smg$ is 
normalized, and hence describes a physical state of $\Gamma$, 
parametrically dependent on $\bm\Omega$. Notice that 
the ${\bm\Omega}$-dependence of $\ket{\phi(\bm\Omega)}$
survives iff $\dket{\Psi}$ is entangled.

There is a certain degree of freedom in the group-theoretic construction 
of GCS (see for instance Tables I and II in Ref.~\cite{ZhangFG90}), due 
to the possibility of choosing different set of generators for 
$\mathfrak{g}_\smc$, 
i.e. different Cartan basis, and an arbitrary state $\ket{G}$ from which to start the construction, so 
called reference state.
As for the generators of semisimple algebras, we remind that the Cartan 
decomposition classifies them into {\it diagonal}, 
$\{\hat D_\delta\}$, and {\it raising} operators, $\{\hat R_m, \hat 
R_{-m}\}$, according to $[\hat D_\delta,\hat D_\theta]=0, 
[\hat D_\delta,\hat R_m]=d_ {\delta m}\hat R_m, 
[\hat R_m,\hat R_{-m}]=\sum_\delta d_{\delta m}\hat D_\delta$, 
and  $[\hat R_m,\hat R_{m'}]=c_{mm'}\hat R_{m+m'}$, where the 
coefficients $\{d_{\delta m}\},\{c_{mm'}\}$ are the so called structure 
constants. 
By way of example, for the semisimple algebras $\mathfrak{su}(2)$ and
$\mathfrak{su}(1,1)$, 
that define the spin and pseudo-spin coherent states, 
respectively, it is $M=1$, with $\hat R_1=\hat S^-$ and $\hat K^-$. 
When spin squeezing is considered, it is $M=2$, with $\hat R_2=(\hat 
S^-)^2$. The non-semisimple algebra ${\mathfrak h}_4$ that defines the 
harmonic-oscillator coherent states has a similar decomposition into 
diagonal ($\hat a^\dagger\hat a, \hat{\mathbb I}$) 
and creation/annihilation ($\hat a^\dagger, \hat a$) operators, leading 
to the same results hereafter derived for the semisimple case, 
as shown in the supplemental material.

We choose the Cartan basis so that $\hat{H}_\smc$ depends linearly on one 
of its diagonal operators only, say $\hat H_\smc=\varsigma\hat D_1+K$, 
where $K$ is a real arbitrary constant and $\varsigma^2=\pm1$ such that 
$\epsilon\defi\varsigma d_{1\ell}$ is real and positive for some $\ell$,
which ensures $\hat H_\smc$ is hermitian.
For the sake of a lighter notation, we also normalize the 
raising and diagonal operators so that 
$\varsigma^2\sum_\delta d^2_{\delta\ell}\to 2$. 
As for the reference $\ket{G}$, we set it as the minimal weight state, 
$\hat R_m\ket{G}=0~\forall m$, which is easily seen to be an eigenstate of the 
diagonal operators, $\hat D_\delta\ket{G}=g_\delta\ket{G}$. In 
particular, hence, it is $\hat H_\smc\ket{G}=\epsilon_0\ket{G}$, with
$\epsilon_0\defi\varsigma g_1+K$, and we will hereafter take $K$ 
so that $\epsilon_0=0$.

Once the Cartan basis and the reference state are chosen, GCS are 
generated via 
\begin{equation}
\ket{\bm \Omega}=
e^{\bm\Omega\cdot\hat{\bm R}^\dagger-\bm\Omega^*\cdot\hat{\bm R}}\ket{G}~, 
\label{e.GCS}
\end{equation} 
where $\hat{\bm R}\defi(\hat R_1,\hat R_2...\hat R_M)$; 
notice that the index $m$ runs from 1 to $M$ both in 
$\Omega_m$ and in $\hat R_m$, by definition. GCS as from 
Eq.~\eqref{e.GCS} are normalized and non-orthogonal, and
expectation values of operators upon them, 
$\exval{\bm\Omega|\hat O|\bm\Omega}$, are often dubbed {\it symbols}, 
indicated by $O({\bm\Omega})$.
For more technical details on this section, we refer the reader to the 
supplemental material.

\section{A quantum clock for a quantum system}
\label{s.letscallit-phi}

We consider the set of GCS defined by 
$\bm\Omega_\ell=(0,0,...,\Omega_\ell,...0)$, 
with $\ell$ chosen at will amongst those for which $\epsilon$ is real 
and positive. Given that $\Omega_\ell\in\mathbb C$, we will hereafter 
use 
\begin{equation}
\lambda\defi\Omega_\ell=\varrho e^{-i\varphi}~,
\label{e.def-Omega_ell}
\end{equation}
with $\varrho\in[0,\infty)$ and $\varphi\in(-\infty,\infty)$.
Using the BCH formulas proper 
to $\mathfrak{g}_\smc$, and the definition \eqref{e.GCS}, it can be easily 
shown that
\begin{equation}
\ket{\lambda}\defi\ket{\bm\Omega_\ell}=
N_{\varrho} e^{\Lambda\hat R^\dagger_\ell}\ket{G}~,
\label{e.ketlambda}
\end{equation}
with $\Lambda=|\tan(\varsigma\varrho)| e^{-i\varphi}$ and
$N_{\varrho}$ a normalization factor that does not depend on $\varphi$.
Furthermore, from the Cartan commutation rule $[\hat D_\delta,\hat R_\ell]
=d_{\delta \ell}\hat R_\ell$ it follows $[\hat H_\smc,e^{\Lambda^*\hat R_\ell}]=
\epsilon\Lambda^*\hat R_\ell e^{\Lambda^*\hat R_\ell}$, leading to
\begin{eqnarray}
&\!&\bra\lambda\hat H_\smc\ket{\bm\Omega}=
\bra{G}N_{\varrho}e^{\Lambda^*\hat R_\ell}\hat H_\smc 
\ket{\bm\Omega}=\nonumber\\
%&=&\bra{G}\left(\hat H_\smc N_{\varrho}e^{\Lambda^*\hat R_\ell}-
%\epsilon\Lambda^*\hat R_\ell N_{\varrho}e^{\Lambda^*\hat R_\ell}\right)
%\ket{\bm\Omega}=\nonumber\\
%&=&\bra{G}\left(\hat H_\smc N_{\varrho}e^{\Lambda^*\hat R_\ell}+
%i\epsilon\frac{d}{d\varphi}N_{\varrho}e^{\Lambda^*\hat R_\ell}\right)
%\ket{\bm\Omega}=\nonumber\\
&=&i\epsilon\frac{d}{d\varphi}
\exval{\lambda|\bm\Omega}~,
\label{e.phi-derivative}
\end{eqnarray}
Once defined the partial inner product 
$\langle\cdot\dket{\cdot}:\HH_\smc\otimes\HH_\smg\rightarrow\HH_\smg$
such that $\bra{\gamma}[\langle\xi\dket{\Psi}]=
(\bra{\gamma}\otimes\bra{\xi})\dket{\Psi}~,\forall 
\xi\in\HH_\smc$ and $\forall \gamma\in{\HH_\smg}$ 
we project the constraint \eqref{e.deWitt} in the form
\begin{equation}
\langle\lambda|\hat H\dket{\Psi}=0~,
\label{e.projected-deWitt}
\end{equation}
with $\hat H$ and $\dket{\Psi}$ as in Eq.~\eqref{e.H} and 
\eqref{e.Psi_para}, and find, by virtue of the result
\eqref{e.phi-derivative},
\begin{equation}
i\epsilon\frac{d}{d\varphi}\ket{\Phi_{\varrho}(\varphi)}
=\hat H_\smg\ket{\Phi_{\varrho}(\varphi)}~,
\label{e.un-Schroedinger}
\end{equation}
where
\begin{equation}
\ket{\Phi_{\varrho}(\varphi)}\defi
\langle\lambda\dket{\Psi}{=}
\int_{{\cal M}_\smc}\!\!d\mu(\bm\Omega)\chi(\bm\Omega)\exval{\lambda|\bm\Omega}
\ket{\phi(\bm\Omega)}
\label{e.un-normState}
\end{equation}
is an un-normalized element of ${\cal H}_\smg$, and we have introduced a 
notation that 
highlights the different meaning that the dependence 
on $\varrho$ will have in what follows, w.r.t. that on 
$\varphi$.
Reminding that $\epsilon$ is real and positive,
Eq.~\eqref{e.un-Schroedinger} has the same 
form of the Shr\"odinger equation, with the real parameter 
\begin{equation}
\frac{\hbar}{\epsilon}\varphi
\label{e.time}
\end{equation}
playing the role of time, as found resorting to other parametric 
representations \cite{PageW83,GiovannettiLM15,BarniBS15,FavalliS20}.
However, Eq.~\eqref{e.un-Schroedinger} is not the Schr\"odinger 
equation, as $\ket{\Phi_\varrho(\varphi)}$ is not normalized.
This is most often considered an amendable fault, as from 
Eq.~\eqref{e.un-Schroedinger} it follows 
$\frac{d}{d\varphi}\exval{\Phi_\varrho(\varphi)|\Phi_\varrho(\varphi)}=0$
meaning that, should $\ket{\Phi_\varrho(\varphi)}$ have a
non-vanishing and finite norm, Eq.~\eqref{e.un-Schroedinger} would also 
hold for its normalized sibling. Before considering this point, let us 
collect some more clues on the meaning of $\varrho$ and $\varphi$.

Getting back to the operator $\hat R_\ell$ introduced at the 
beginning of this section, one can define~\cite{CarruthersN65,VaglicaV84}
the so called "phase-operator" $\hat{\phi}$, 
via
\begin{equation}
\hat{R}_\ell=(\hat{R}_\ell\hat{R}_\ell^\dagger)^{1/2}e^{-i\hat{\phi}}~. 
\label{e.phase-operator} 
\end{equation} 
From the commutation rules between elements of the Cartan basis, 
reminding that $\hat H_\smc=\varsigma\hat D_1+K$ and 
$\epsilon=\varsigma d_{1\ell}\in\mathbb{R}^{+}$, 
it follows
\begin{equation}
[\hat{H}_\smc,\sin\hat{\phi}]=i\epsilon\cos\hat{\phi}~, 
\label{e.phase-commutation}
\end{equation} 
and hence (see for instance Ref.~\cite{BransdenJ00}) 
\begin{equation}
\Delta \hat{H}_\smc\Delta\sin\hat{\phi} \geq
\left|\frac{\epsilon}{2}\exval{\cos\hat{\phi}}\right|~, 
\label{e.uncertainty-sinphi}
\end{equation} 
with 
$\Delta\hat{B}
\defi(\langle \hat{B}^2\rangle-\langle\hat{B}\rangle^2)^{1/2}$ for any 
hermitian operator $\hat B$.
Noticing that Eqs.~\eqref{e.deWitt}-\eqref{e.H} imply 
a relation between $\hat H_\smc$ and the energy of the 
system, while Eqs.~\eqref{e.def-Omega_ell} 
and \eqref{e.phase-operator} 
relate $\hat\phi$ with $\varphi$, 
one might say that the inequality \eqref{e.uncertainty-sinphi} 
is the ancestor of the time-energy uncertainty relation for $\Gamma$, 
after setting $\varphi\ll1$ and the parameter \eqref{e.time} precisely as time, 
a statement that is made clear in the next section.

Summarizing, we have so far collected results that point to 
$\hbar\varphi/\epsilon$ as "the time" for the evolving system, but
the overall picture is not that provided by QM, where the 
quantum character of the clock is totally absent; this is the reason why 
we take our next step.

\section{A classical clock for a quantum system}
\label{s.letscallit-t}

We now assume that the quantum theory describing $C$ satisfies the 
conditions ensuring it flows into a well defined classical theory
when the clock becomes macroscopic, according to the large-$N$ 
quantum approach based on GCS, 
as briefly described in the Introduction.
In particular, we use that GCS are the only quantum states 
that survive the quantum-to-classical crossover, 
insofar doing becoming orthogonal 
\begin{equation}
\lim_{N\to\infty}
\exval{\bm\Omega|\bm\Omega'}\to\delta(\bm\Omega-\bm\Omega')~,
\label{e.GCS-orthogonal}
\end{equation}
and defining the classical states identified by the 
corresponding points $\bm\Omega$ on the classical phase-space ${\cal 
M}$.
As for the observables, the only ones that stay meaningful throughout 
the crossover must obey
\begin{equation}
\lim_{N\to\infty}
\frac{\exval{\bm\Omega|\hat A|\bm\Omega'}}
{\exval{\bm\Omega|\bm\Omega'}}<\infty~,
\label{e.symbols}
\end{equation}
so as to transform into well defined functions on the classical 
phase-space.
Using Eq.~\eqref{e.GCS-orthogonal} one can easily show that 
$\exval{\Phi_\varrho(\varphi)|\Phi_\varrho(\varphi)}\to
\chi^2(\lambda)$ in the classical limit for the clock;
moreover, it is $\chi^2(\lambda)\equiv\chi^2(\varrho)$ due 
to Eq.~\eqref{e.un-Schroedinger}. Therefore, reminding that
$\chi^2(\varrho)$ is a normalized probability distribution, 
any $\varrho$ for which $\chi^2(\varrho)\neq 0$ defines a
physical state
\begin{equation} 
\ket{\phi_\varrho(\varphi)}\defi 
\frac{\ket{\Phi_\varrho(\varphi)}}{\sqrt{\chi^2(\varrho)}}~,
\label{e.parametric-state} 
\end{equation}
whose dependence on $\varphi$ is ruled by
\begin{equation} 
i\epsilon\frac{d}{d\varphi}\ket{\phi_{\varrho}(\varphi)} =
\hat H_\smg\ket{\phi_{\varrho}(\varphi)}~, 
\label{e.Schroedinger} 
\end{equation} 
which is the Schr\"odinger equation with $t=\hbar\varphi/\epsilon$.
In fact, the above result is a derivation of the Schr\"odinger 
equation akin to that suggested in the original work by Page and 
Wootters~\cite{PageW83}, with state-normalization ensured by 
construction, for a classical clock. We notice, though,
that as a byproduct of having specifically addressed the normalization issue, 
the state \eqref{e.parametric-state} has a further dependence on the 
real parameter $\varrho$.
In order to understand its meaning as far as the evolving system is 
concerned, we get back to the constraint 
\eqref{e.deWitt} and its projection upon a GCS $\ket{\lambda}$ of the 
clock, Eq.~\eqref{e.projected-deWitt}, with $\dket{\Psi}$ as in 
Eq.~\eqref{e.Psi_para}. Considering that $\exval{\lambda|\bm\Omega}$ is 
finite for finite $N$, we write
\begin{eqnarray}
\!\!\!&0&=\bra{\lambda}\hat H\dket{\Psi}=\nonumber\\
\!\!\!&=&\!\!\int_{\cal 
M}\!\!\!d\mu(\bm\Omega)\chi(\bm\Omega)\exval{\lambda|\bm\Omega}
\left(\frac{\exval{\lambda|\hat H_\smc|\bm\Omega}}{\exval{\lambda|\bm\Omega}}-
\hat H_\smg\right)\ket{\phi(\bm\Omega)}~
\label{e.parametric-deWitt}
\end{eqnarray}
that becomes, in the classical limit for $C$
where Eqs.~\eqref{e.GCS-orthogonal} and \eqref{e.symbols} hold
and for any $\varrho$ such that $\chi^2(\varrho)\neq 0$,
\begin{equation}
\hat H_\smg\ket{\phi_\varrho(\varphi)}=
E_\smg(\varrho)\ket{\phi_\varrho(\varphi)}~,
\label{e.stationary-Schroedinger}
\end{equation}
with
\begin{equation}
E_\smg(\varrho)=\exval{\lambda|\hat H_\smc|\lambda}~;
\label{e.E_varrho}
\end{equation} 
the r.h.s. of the above equation, which is the symbol of $\hat H_\smc$ on 
$\ket{\lambda}$, can be calculated and reads (see supplemental 
material)
\begin{equation}
H_\smc(\varrho)\defi\exval{\lambda|\hat H_\smc|\lambda}=
\frac{\epsilon}{2}b^2\left(\cos(2\varsigma\varrho)-1\right)~,
\label{e.HC-symbol}
\end{equation}
with $\varsigma^2 b^2=\sum_\delta g_\delta d_{\delta\ell}$. It is 
relevant that 
Eq.~\eqref{e.HC-symbol} follows from algebraic properties, 
and therefore holds in general, regardless of the details of the theory 
that describes the clock. Furthermore, $H_\smc(\varrho)$ does not depend on 
$\varphi$, 
which justifies the use of the notation $E_\smg(\varrho)$ in 
Eq.~\eqref{e.E_varrho} and allows one to consistently relate 
Eq.~\eqref{e.stationary-Schroedinger} with the stationary 
Schr\"odinger equation for $\Gamma$, with $\varrho$ the
parameter that sets its energy.

\subsection*{An uncertainty relation}
\label{ss.uncertainty}

Let us now consider what happens when making measurements on the clock.
We know that GCS are the only quantum states that 
survive the quantum-to-classical crossover according to 
$\ket{\bm\Omega}\to{\bm\Omega}$, as described above and thoroughly 
discussed in the literature \cite{Zurek03,LiuzzoScorpoCV15epl,RossiEtal17,FotiEtal19,Foti_Phd19}. 
This means that performing a quantum measurement upon a system whose 
behaviour can be effectively described as if it were classical,  
is tantamount to select one GCS $\ket{\bm\Omega}$ to be the ancestor of 
the observed classical state or, which is the same,
say that the combined effect of a  measurement and the classical limit 
is to make $\chi^2({\bm\Omega})$ 
become a Dirac-$\delta$ around the point $\bm\Omega$ on ${\cal M}_\smc$ 
that identifies the observed classical state.
Let us now take such state to be one of the GCS $\ket{\lambda}$, 
consistently with the task of making measurements of observables 
that characterize it as a clock, such as $\hat H_\smc$ or 
$\sin\hat\phi$ in Eq.~\eqref{e.phase-commutation}. 
When taking the classical limit of the clock, it can be 
demonstrated~\cite{CarruthersN65,VaglicaV84} that
\begin{equation}
\exval{\lambda|\sin\hat\phi|\lambda}\to\sin\varphi~~,~~
\exval{\lambda|\cos\hat\phi|\lambda}\to\cos\varphi~;
\label{e.exval-sincos_phi}
\end{equation}
this result, together with the definition
$\Delta E_\smg(\varrho)\defi\Delta H_\smc(\varrho)$ (that follows from 
Eqs.~\eqref{e.stationary-Schroedinger}-\eqref{e.E_varrho})
and a  small-$\varphi$ approximation, provides 
\begin{equation}
\Delta E_\smg(\varrho)\Delta\varphi\ge\frac{\epsilon}{2}~,
\label{e.uncertainty-for-Gamma}
\end{equation}
which we recognize, once the parameter $\hbar\varphi/\epsilon$ is 
identified with time, as a proper energy-time uncertainty relation for 
$\Gamma$. We will further comment upon this result in the 
concluding section.

Collecting all the clues so far obtained, we conclude this section 
mantaining that the parameter 
\eqref{e.time}
is what we call "time" in QM, a statement that we express as
\begin{equation}
t^{\rm QM}=\frac{\hbar}{\epsilon}\varphi~,
\label{e.time-QM}
\end{equation}
where the apex QM indicates that this is the parameter that enters the 
quantum description of evolving systems.

This is not the end of the story, though, because it is now necessary to 
demonstrate that when the system $\Gamma$ undergoes the 
quantum-to-classical crossover, the above results lead to the  
Hamilton e.o.m., with the parameter $\hbar\varphi/\epsilon$ still 
playing the role of time. To this purpose, in the next section we take 
the classical limit also for the evolving system, thus moving into a 
completely classical setting.

\section{A classical clock for a classical system}
\label{s.emergence}

Let us now consider what happens when the system $\Gamma$ 
becomes macroscopic in a way that makes its behaviour amenable to the 
laws of classical physics. As in the previous section, the problem is 
tackled in terms of GCS in the large-$N$ limit.
Therefore, besides the GCS for the clock $\{\ket{\bm\Omega}\}$ defined 
in 
Sec.~\ref{s.non_interacting}, here we also use the GCS for the 
system, i.e. those relative to the Lie algebra 
$\mathfrak{g}_\smg$ proper to the quantum 
theory that describes $\Gamma$. These will be indicated by 
$\{\ket{\bm\gamma}\}$, where ${\bm\gamma}=(\gamma_1,\gamma_2,...\gamma_J)$ 
with $\gamma_j\in\mathbb C~\forall j$, and $J$ related to the dimension 
of $\mathfrak{g}_\smg$. Each $\ket{\bm\gamma}$ univocally identifies one 
point on the manifold ${\cal M}_\smg$, whose (real) dimension 
is $2J$. 

Using the resolution of the identiy upon ${\cal H}_\smc$ and ${\cal 
H}_\smg$ in terms of the GCS $\{\ket{\bm\Omega}\}$ and $\{\ket{\bm\gamma\}}$, 
respectively, we write the state $\dket{\Psi}$ of the overall system as
\begin{equation}
\dket{\Psi}=
\int_{{\cal M}_\smc} \!\!\!d\mu(\bm\Omega)
\int_{{\cal M}_\smg}\!\!\! d\mu(\bm\gamma)
\beta(\bm\Omega,\bm\gamma)\ket{\bm\Omega}\otimes\ket{\bm\gamma}~,
\label{e.Psi_doublepara}
\end{equation}
where 
\begin{equation}
\beta(\bm\Omega,\bm\gamma)\defi
(\bra{\bm\Omega}\otimes\bra{\bm\gamma})\dket{\Psi}
=\chi(\bm\Omega)\langle\bm\gamma\ket{\phi(\bm\Omega)}
\label{e.beta}
\end{equation}
is a function on ${\cal M}_\smc\times{\cal M}_\smg$ whose square modulus,
$\chi^2({\bm\Omega})|\exval{{\bm\gamma}|\phi({\bm\Omega})}|^2$
is the conditional probability for $\Gamma$ to be in the 
state $\ket{\bm\gamma}$ when $C$ is in the state $\ket{\bm\Omega}$, 
given that the global system $\Psi$ is in the pure state $\dket{\Psi}$. 
In other terms, $\beta({\bm\Omega},{\bm\gamma})$ is different from zero 
only on those pairs 
$({\bm\Omega},{\bm\gamma})\in{\cal M}_\smc\times{\cal M}_\smg$ that
define states 
$\ket{\bm\Omega}\otimes\ket{\bm\gamma}\in{\cal H}_\smc\otimes{\cal H}_\smg$ 
which are present in the decomposition of $\dket{\Psi}$ in terms of GCS, 
Eq.~\eqref{e.Psi_doublepara}.

Projecting the constraint \eqref{e.deWitt} 
upon one specific state 
$\ket{\overline{\bm\Omega}}\otimes\ket{\overline{\bm\gamma}}$, we write
\begin{eqnarray}
0&=&\bra{\overline{\bm\Omega}}\otimes\bra{\overline{\bm\gamma}}
\hat H\dket{\Psi}=\nonumber\\
&=&\int_{{\cal M}_\smc} \!\!\!d\mu(\bm\Omega)
\int_{{\cal M}_\smg}\!\!\! d\mu(\bm\gamma)\beta(\bm\Omega,\bm\gamma)
\exval{\overline{\bm\Omega}|\bm\Omega}
\exval{\overline{\bm{\gamma}}|\bm\gamma}\times\nonumber\\
&~&\times\left[
\frac{\exval{\overline{\bm\Omega}|\hat H_\smc|\bm\Omega}}{\exval{\overline{\bm\Omega}|\bm\Omega}}
-
\frac{\exval{\overline{\bm\gamma}|\hat H_\smg|\bm\gamma}}{\exval{\overline{\bm\gamma}|\bm\gamma}}
\right]
\end{eqnarray}
that becomes, in the classical limit for $C$ and $\Gamma$,
i.e. assuming Eqs.~\eqref{e.GCS-orthogonal} and \eqref{e.symbols} hold 
not only for the GCS and the Hamiltonian 
of the clock but also for those of the system,
\begin{equation}
H_\smc(\bm\Omega)=H_\smg(\bm\gamma)
\label{e.deWitt-classical}
\end{equation}
for $({\bm\Omega},{\bm\gamma})$ such 
that $\beta({\bm\Omega},{\bm\gamma})\neq 0$,
meaning that the configurations $({\bm\Omega},{\bm\gamma})$ into which 
the original quantum state $\dket{\Psi}$ can flow when clock 
and system behave according to the rules of classical 
physics, must obey Eq.~\eqref{e.deWitt-classical}.
In particular, if one considers the configurations amongst those 
for which $\beta({\bm\Omega},{\bm\gamma})\neq 0$ that 
have ${\bm\Omega}=(0,0,...\Omega_\ell,...0)$,
corresponding to the GCS $\ket{\lambda}$  introduced in 
Sec.~\ref{s.letscallit-phi} and identified by the 
complex variable $\lambda=\varrho e^{-i\varphi}$, these will belong
to a 
submanifold  $({\cal U}_\smc\subset\mathbb{C})\times({\cal 
U}_\smg\subset{\cal M}_\smg$) such that a map $F:{\cal U}_\smc\to{\cal 
U}_\smg$ exists, defined by
\begin{equation}
\lambda\in{\cal U}_\smc\underset{F}{\longrightarrow}{\bm u}\in{\cal U}_\smg
~:~
H_\smg(\bm u=F(\lambda))=H_\smc(\varrho)~.
\label{e.the-map}
\end{equation}
As the explicit form of $F$ is arbitrary, we fix it as follows.
We consider that ${\cal M}_\smg$ has a 
symplectic structure, which means that it exists a Darboux chart 
%\cite{???} 
\begin{equation}
\!\!\begin{dcases}
{\cal D}:\bm\gamma\in{\cal M}_\smg\!\!\rightarrow
({\bm q},{\bm p})\!
\defi\!((q_1,p_1),(q_2,p_2)...(q_J,p_J))\in\mathbb{R}^{2J}\!,\\
\mbox{such that } \{q_i,p_j\}_\smg=
\mathfrak{h}^{-1}\delta_{ij}~~{\rm with}~~\mathfrak{h}=const.~,\\
\mbox{where}\;\;
\{\cdot,\cdot\}_\smg \;\mbox{are Poisson brackets on\;} 
\mathcal{M}_\smg~,
\end{dcases}
\label{e.Darboux-chart}
\end{equation}
that relates the parametrization of GCS via $J$-dimensional 
complex vectors $\{\bm\gamma\}$ with that obtained via $J$ pairs of 
real, canonically conjugated, variables $(q_j,p_j)$.
For these pairs, referring to Ref.~\cite{ZhangFG90}, we choose
\begin{equation}
q_j-i\varsigma^2 
p_j=v_j\sqrt{2}b\varsigma\sin(\varsigma\varrho)e^{-i\varphi}
\label{e.explicit-F}
\end{equation}
with $\vec{v}\in\mathbb{R}^J$ constant unit vector, i.e. $\sum_{j} 
v_j^2=1$. As far as condition \eqref{e.the-map} is fulfilled, other choices are possible, without affecting 
the overall scheme and the subsequent results.
Once $F$ is given, the so called "pullback-by-$F$" 
map, sometimes indicated by $F^*$, is also defined, according to
$F^*:\omega^{(k)}_\smg\longrightarrow\omega^{(k)}_{\smc}$,
where $\omega^{(k)}_{\smg(\smc)}$ are 
$k$-forms on ${\cal U}_{\smg(\smc)}$. In particular, for $k=0$, 
i.e. when considering functions, it is 
$(F^*f_\smg)(\lambda)=f_\smc(\lambda)$, with 
$F^*f_\smg=f_\smg({\bm u}=F(\lambda))$.
Applying $F^*$ on the symplectic 2-form defining the 
standard Poisson brackets in \eqref{e.Darboux-chart}, 
we obtain the Poisson brackets on ${\cal M}_\smc$, 
that read (see supplemental material)
\begin{align}
&\{f_\smc,g_\smc\}_{\smc}=
\frac{1}
{\mathfrak{h}b^2\varsigma
\sin\left(2\varsigma\varrho\right)}\left(
\frac{\partial f_\smc}{\partial\varrho}
\frac{\partial g_\smc}{\partial\varphi}-
\frac{\partial g_\smc}{\partial\varrho}
\frac{\partial f_\smc}{\partial\varphi}\right)
\label{e.PB_clock}
\end{align} 
$\forall\;f_\smc,g_\smc$ generic functions on $\mathcal{M}_\smc$.
On the other hand,  $q_j$ and $p_j$ are by all means functions on 
$\mathcal{M}_\smc$, as seen in Eq.~\eqref{e.explicit-F}; therefore, 
using Eq.~\eqref{e.PB_clock} with $g_\smc=H_\smc(\varrho)$ from 
Eq.~\eqref{e.HC-symbol} we evaluate
$\{q_j,H_\smc\}_\smc$ and $\{p_j,H_\smc\}_\smc$, and find (see 
supplemental material)
$\{q_j,H_\smc\}_\smc
=\frac{\epsilon}{\mathfrak{h}}\frac{d\,q_j}{d\,\varphi}$, 
and 
$\{p_j,H_\smc\}_\smc
=\frac{\epsilon}{\mathfrak{h}}\frac{d\,p_j}{d\,\varphi}$.
Finally, using 
$
\{f_\smc(\lambda),g_\smc(\lambda)\}_\smc=
%\{f_\smg(\bm u =F(\lambda),g_\smg(\bm u =F\lambda)\}_\smc=
\{f_\smg(\bm u),g_\smg(\bm u)\}_\smg~,
$ we obtain
\begin{equation}
\begin{dcases} 
\{q_j,H_\smg\}_\smg=\frac{\epsilon}{\mathfrak{h}}\frac{d\,q_j}{d\,\varphi}\\
\{p_j,H_\smg\}_\smg=\frac{\epsilon}{\mathfrak{h}}\frac{d\,p_j}{d\,\varphi} 
\end{dcases} 
\label{e.Hamilton-eom}
\end{equation} 
i.e. the Hamilton e.o.m. ruling the dynamics of classical systems, once 
time is recognized as the parameter  
\begin{equation}
t^{\rm CL}=\frac{\mathfrak{h}}{\epsilon}\varphi~,
\label{e.classical-time}
\end{equation}
where the apex CL indicates that this is the parameter that
enters the classical description of evolving systems.
Getting back to Eq.~\eqref{e.time-QM} and setting the arbitrary constant 
$\mathfrak{h}$ in the Poisson-brackets of the Darboux chart 
\eqref{e.Darboux-chart} equal to $\hbar$, we finally obtain
\begin{equation}
t^{\rm QM}=t^{\rm CL}=\frac{\hbar}{\epsilon}\varphi~.
\label{e.onetime}
\end{equation}
This last equation, together with the 
derivation in one same framework of both the 
quantum-mechanical Schr\"odinger equation \eqref{e.Schroedinger} 
and the classical Hamilton e.o.m. \eqref{e.Hamilton-eom},
represents the main result of this work, which is discussed 
in the next and last section. 

\section{Discussion, conclusions, and further developments}
\label{s.conc-disc}

In the last decades we have learnt that when quantum macroscopic
systems can be effectively studied as if they were classical (which is 
what should be meant by "classical"), their geometrical properties follow 
from the algebraic structure of the quantum theory originally 
describing them (see for instance the way a specific phase-space emerges 
as the symplectic manifold involved in the GCS construction for one 
assigned quantum Lie-algebra). This is by itself quite a breakthrough, 
as it allows to establish a dialogue between classical and quantum 
physics without resorting to disjointed interventions such as 
quantization or, in the opposite direction, non-unitary state-reduction.

When considering more than one system, things become ever more 
interesting. In fact, when a quantum system interacts with a 
classical environment (be that a magnetic field, or a 
thermal bath, or some macroscopic environment), the pure states of the 
former acquire a parametric dependence that testifies the existence of 
the latter, and gives rise to geometrical effects such as the quantum 
Berry-phase \cite{Vedral03,Nori08,CCGV13pnas,RotondoN19}.
Awe comes, though, as these effects emerge even without interaction,
as far as the systems are entangled and some physical constraint is 
enforced, such as Eq.~\eqref{e.deWitt} in the PaW mechanism. 
Indeed, this is how
states of a quantum system come to depend on time according to the 
Schr\"odinger equation, as also shown in this work. 
In such setting, coordinates of points in manifolds and 
elements of Hilbert spaces (e.g.
$\varrho,\varphi$ and $\ket{\phi_\varrho(\varphi)}$
in this work) relate to each 
other via rules, such as the Schr\"odinger equation or the 
time-energy uncertainty relation, whose 
generality is that of the physical principles. 
To this respect we like to comment upon two of our results:
First we notice that the energy of the system $\Gamma$, i.e. 
$E_\smg(\varrho)$ in Eq.~\eqref{e.stationary-Schroedinger}, 
does not depend on time, i.e. on $\varphi$, consistently with the fact 
that the Hamiltonian of an isolated system cannot depend on time. 
Then we underline that, as in Refs.~\cite{FavalliS20,MacconeS20}, 
the inequality \eqref{e.uncertainty-for-Gamma}
does not follow from the non-commutativity between $\hat H_\smg$
and some other operator acting on ${\cal H}_\smg$:
it is rather an indirect consequence of the inequality 
\eqref{e.phase-commutation}, which regards operators acting on the 
clock, plus the constraint \eqref{e.deWitt} and the possibility, given 
by the use of GCS, of describing the clock as a classical object without 
wiping out one of its most relevant quantum feature, namely its being 
entangled with the evolving system.

What is most remarkable, though, is that a genuinely quantum feature 
such as entanglement survives even in a completely classical setting,
there continuing to cause the emergence of such a fundamental ingredient 
of our everyday life as time, which is what we have here 
demonstrated by deriving the Hamilton e.o.m~\eqref{e.Hamilton-eom}.
In fact, our results in the fully classical setting unravel another tangle of classical physics, namely the 
relation between phase-space and space-time.
This relation emerges from the fact that when the global system is in 
the pure state $\dket{\Psi}$, the only configurations that 
survive its classical limit are those identified by points 
$(\bm\Omega,\bm\gamma)\in{\cal M}_\smc\times{\cal M}_\smg$ where the 
probability $|\beta(\bm\Omega,\bm\gamma)|^2$ is different from zero.
Therefore, while the phase-space of $\Gamma$ is the $2J$ dimensional 
simplectic manifold ${\cal M}_\smg$ defined by the
GCS $\ket{\bm\gamma}$ introduced in Sec.~\ref{s.emergence}, its
space-time is the $(J+1)$-dimensional real hypersurface defined by 
Eqs.~\eqref{e.deWitt-classical} and \eqref{e.explicit-F}, whose points 
(i.e. events) are identified by the coordinates 
$(\hbar\varphi/\epsilon;\bm q)$, with 
$\varphi=-\arg\lambda\in\mathbb{R}$ from Eq.~\eqref{e.def-Omega_ell} and 
$\bm q=\bm q(\bm\gamma)\in\mathbb{R}^J$ from the Darboux chart 
\eqref{e.Darboux-chart},
such that 
$\beta(\overline\varrho,\varphi;\bm q,\overline{\bm p})$ 
is different from zero for some  $\overline\varrho$ (i.e. energy of the 
clock) and $\overline{\bm p}$ (i.e. momentum of the system).
Notice that, if $C$ and $\Gamma$ were not
entangled, i.e $\dket{\Psi}=\ket{C}\otimes\ket{\Gamma}$, it would be 
$\beta(\overline\varrho,\varphi;{\bm q},\overline{\bm p})=
\chi(\overline\varrho) 
\exval{(\bm{q},\overline{\bm p})|\Gamma}$, 
with no relation between instants of time $\varphi$
and position in space ${\bm q}$, i.e. with no causal relation between 
events. In other terms, as emerged in different contexts 
(see for instance Ref.~\cite{VanRaamsdonk10}) not only quantum 
entanglement is what makes physical 
systems to evolve, but it also provides their spacetime with a causal 
structure. 

Despite effects of entanglement without interaction being already phenomenal, 
we think that taking possible interactions into account will lead to 
substantial developments of this work.
One might first consider adding a quantum environment
with which $\Gamma$ starts interacting while being already entangled 
with the clock. This should describe the dynamics of the density 
operator of $\Gamma$, 
and show how, and under what conditions, the Liouville-VonNeumann 
equation emerges, with clues about the non-unitary evolution of 
non-isolated systems.
The presence of multiple clocks, possibly interacting amongst 
themselves, also seems an intriguing enrichment, particularly in view 
of some recent works by other authors~\cite{BryanM18,SmithA19,CastroRuizEtal20}.
However, the most compelling follow-up of this work, in our opinion, is that of
relating the picture it proposes with that provided by relativity.
In fact, we expect relativistic quantum mechanics and 
quantum-field-theory to find their place in the hybrid setting of 
Sec.~\ref{s.letscallit-t}, where studying how the
expectation values of operators on ${\cal H}_\smg$ get to depend on 
$(\varrho,\varphi)$ via the parametric dependence
of the states $\ket{\phi_\varrho(\varphi)}$, might help understanding 
some 
unclear aspects of the way special relativity encounters quantum mechanics.
Moreover, having connected the classical formalism that set the scene
for general relativity and gravity with a full quantum description, 
we think we have ideal tools for breaking through some of the obstacles 
that make quantum gravity so difficult to process.
In particular, we believe that studying the probability distribution 
$|\beta(\lambda;{\bm q},{\bm p})|^2$ in relation to the original Lie 
algebras $\mathfrak{g}_\smc$ and $\mathfrak{g}_\smg$ and/or the specific 
form of the quantum Hamiltonian $\hat H$ may provide a link between the 
geodesic principle and the Schr\"odinger equation; furthermore, taking 
into account a possible interaction between evolving system and clock, 
as suggested in Ref.~\cite{SmithA19}, or between different clocks, as in 
Ref.~\cite{CastroRuizEtal20}, might explain spacetime 
deformation, and hence gravity, from a quantum viewpoint. Work in this 
direction is in progress, particularly referring to the case of 
Schwartzschild black-holes \cite{CoppoMS19} and Hawking radiation \cite{PranziniMS20}.\\

\section{Acknowledgements}
We thank G.~Garc\'ia-P\'erez, S~.Maniscalco, L.~Maccone, B.~Sokolov 
and F. Bonechi for amazing discussions, stimulating questions, and valuable 
suggestions. CF and PV also acknowledge the beauty of the Dolomites for 
inspiration during the {\it Quantum Hiking 2019} international 
conference. 
PV gratefully acknowledges support from the Kavli Institute for 
Theoretical Physics @UC Santa Barbara, in the framework of the program
{\it Open Quantum System Dynamics}, and all the discussions with the participants during 
the program itself. AC and CF acknowledge Fondazione CR Firenze for
financial support within project 2018.0951. This work is done in
the framework of the Convenzione operativa between the Institute for
Complex Systems of the Consiglio Nazionale delle Ricerche (Italy) and
the Physics and Astronomy Department of the University of Florence.

%\bibliographystyle{unsrtnat}
%\bibliographystyle{unsrt}
%\bibliography{CateBiblioBis}

%\newpage

\section{Supplementary material}
\subsection{GCS: General Coherent States} 
\label{s_insup-GCS}
Generalized Coherent States (GCS) are an extension of
the field coherent states firstly introduced by R. Glauber in $1963$ 
\cite{Glauber63b}. 
The group-theoretic construction was derived ten years later 
by A. Perelomov \cite{Perelomov72} and R. Gilmore \cite{Gilmore72}, 
independently.
GCS are normalized elements of Hilbert spaces which are in 
one-to-one correspondence with the points of a smooth manifold, 
that has all the properties requested to a classical phase-space. 
In the following, we briefly introduce GCS
according to the procedure described by Gilmore and coworkers in 
Ref.~\cite{ZhangFG90}. 

In order to construct GCS, three inputs are necessary:

\noindent {\it i1)} a Lie-algebra $\mathfrak{g}$, or the related Lie-group ${\cal G}$, 

\noindent {\it i2)} a Hilbert space $\HH$ which is the carrier space of an irreducible 
representation of $\mathfrak{g}$, and 

\noindent {\it i3)} a normalized element $\ket{G}$ of $\HH$. 

Referring to a specific system 
for which GCS are to be constructed, the inputs are as follows:
$\HH$ is the Hilbert space of the system;
$\mathfrak{g}$ is the Lie-algebra whose representation via operators on $\HH$ 
contains the Hamiltonians of the system, meaning that the 
representation of the related Lie-Group ${\cal G}$ contains
all its propagators, which is why ${\cal G}$ is often
dubbed {\it dynamical} group. The normalized element $\ket{G}$ of $\HH$ 
is a physically accessible state of the system, usually called 
{\it reference} state. For the sake of clarity we will hereafter 
identify $\mathfrak{g}$ and ${\cal G}$ with their 
respective representations on $\HH$.
Once the inputs are given, the procedure returns three outputs:

\noindent {\it o1)} the subgroup 
${\cal F}\subset \cal{G}$ whose elements leave $\ket{G}$ 
unchanged apart from an irrelevant overall phase,
and the associated coset $\cal G/\cal F$, such that
every $\hat {\mathbf g}\in\cal G$ can be written as a unique decomposition of two 
group-elements, one belonging to $\cal F$ and the other to $\cal G/\cal 
F$, i.e. $\hat{\mathbf g}=\hat{\mathbf\Omega}\hat{\mathbf{f}}$ with 
$\hat{\mathbf g}\in{\cal{G}},
\,\hat{\mathbf f}\in{\cal{F}},\,\hat{\mathbf\Omega} \in\cal{G}/\cal{F}$;

\noindent {\it o2)} the GCS
\begin{equation}\label{e.GCS_def_inapp}
\ket{\mathbf\Omega}:=\hat{\mathbf\Omega}\ket G\,~,~\forall 
\hat{\bm\Omega}\in{\cal G}/{\cal F}~;
\end{equation}

\noindent{\it o3)} a measure $d\mu(\hat{\mathbf\Omega})$ on ${\cal G}/{\cal F}$ which 
is invariant under the action of the elements of ${\cal G}$, and 
therefore dubbed {\it invariant measure}, such that a resolution of the 
identity upon $\HH$ is provided
\begin{equation}\label{e.id_resolution_inapp}
\int_{\cal{G/F}}
d\mu(\hat{\mathbf\Omega})\ket{\mathbf\Omega}\bra{\mathbf\Omega}=\Id_\HH\,.
\end{equation}
The GCS are normalized,
$\scalar{\mathbf \Omega}{\mathbf \Omega}
=\Bra{G}\hat{\mathbf g}^{-1}\hat{\mathbf g}\Ket{G}= \scalar{G}{G}=1$, 
$\forall\,\hat{\mathbf g}\in \mathcal{G}$, but non-orthogonal,
\begin{eqnarray*}
&~&\scalar{\mathbf \Omega}{\mathbf \Omega'}=
\Bra{G}\hat{\mathbf \Omega}^{-1}\hat{\mathbf \Omega}'\Ket{G}=\\
&~&=\Bra{G}\hat{\mathbf g}^{-1}\hat{\mathbf g}'\Ket{G}e^{i\theta}=
\Bra{G}\hat{\mathbf g}''\Ket{G}e^{i\theta}\neq 0~,
\end{eqnarray*}
$\forall\;\hat{\mathbf g},\hat{\mathbf g}',\hat{\mathbf g}'' \in 
\mathcal{G}$, and $\hat{\mathbf \Omega},\hat{\mathbf \Omega}'\in 
\mathcal{G}/\mathcal{F}$.
For this reason they are said to provide an "overcomplete" set of states 
for ${\cal H}$, where "complete" refers to
Eq.~\eqref{e.id_resolution_inapp}, while "over" means that they are too
many for being all orthogonal to each other.

As for the reference state $\ket{G}$, a common, yet not mandatory, 
choice is that of taking it as an extremal state; for instance, one can 
choose $\ket G$ as the minimal-weight state such that $\hat R_m\ket 
G=0~\forall m$, with $\hat R_m$ defined below.

Getting an explicit expression for the operators 
$\hat{\bm\Omega}$, and hence of the GCS via Eq.~\eqref{e.GCS_def_inapp}, 
requires a characterization of the algebra. 
In particular, if $\mathfrak{g}$ is semisimple, one can  consider its 
Cartan decomposition, that classifies the 
generators as {\it diagonal}, 
$\{\hat D_\delta\}$, or {\it raising}, $\{\hat R_m\,,\,\hat 
R_{-m}\}$, operators, according to 
\begin{eqnarray}\label{e.GCS_Cartan_inapp}
\mbox{[}\hat D_\delta,\hat{D}_\theta\mbox{]}=0\quad &,&
\quad\mbox{[}\hat{D}_\delta,\hat R_m\mbox{]}=d_{\delta m}\hat{R}_m\,,\nonumber\\
\mbox{[}\hat{R}_m,\hat{R}_{-m}\mbox{]}=\sum_\delta d_{\delta m}\hat{D} _\delta\quad &,&\quad
\mbox{[}\hat{R}_m,\hat{R}_{m'}\mbox{]}=c_{m m'}\hat{R} _{m+ m'}\;.\nonumber\\
\end{eqnarray}
where $\{d_{\delta m}\}$, $\{c_{m m'}\}$ are the so called structure 
constants, while $m,m'$ and $\delta,\theta$ go from 1 to some upper 
value $M$ and 
$D$, respectively, that depend on the algebra itself (in the case of 
$\mathfrak{su}(2)$, for instance, it is $M=D=1$, and if spin-squeezing 
is also considered, it is $M=2$ and $D=1$).
In any irreducible representation of $\mathfrak{g}$ it is possible to 
choose the raising operators such that 
$\hat{R}_m^{\dagger}=\hat{R}_{-m}~\forall m$, and, consistently, 
hermitian or anti-hermitian diagonal operators 
$\hat{D}_\delta^{\dagger}=+(-)\hat{D}_\delta\quad\forall\,\delta$, 
depending on the structure constants $\lbrace d_{\delta m}\rbrace$ being 
real or imaginary. The diagonal operators have the reference state 
amongst
their eigenstates, i.e., $\hat D_\delta\ket G=g_{\delta}\ket G$
$\forall \delta$.
Once the Cartan decomposition is available, it can be shown that the 
elements of ${\cal G}/{\cal F}$ in the definition 
\eqref{e.GCS_def_inapp} take the form
\begin{equation}\label{e.GCS_displ_inapp}
\hat{\mathbf\Omega}=\exp{\left(\sum_m\Omega_m\hat R_m^\dagger -
\Omega_m^*\hat R_m\right)}\;,
\end{equation}
where the coefficients $\Omega_m\in\mathbb{C}$ are
coordinates of one point $\mathbf{\Omega}$ of the 
differentiable manifold $\mathcal{M}$, which is associated to 
$\mathcal{G}/\mathcal{F}$ via the {\it quotient manifold} 
theorem~\cite{Lee12}. 
Using a complex projective representation of $\mathcal{G}/\mathcal{F}$,
GCS can also be written as
\begin{equation}
\ket{\mathbf\Omega}=N(|\mathbf{\eta}(\mathbf\Omega)|)\,
e^{\sum_{m}{\eta_m \hat{R}_m^\dagger}}\ket{G}
\label{e.unnorm-GCS_inapp}
\end{equation}
where the normalization constant $N(|\mathbf\eta(\mathbf\Omega)|)$ 
and the relation between the $\eta_m$-coordinates and the $\Omega_m$ ones 
can be obtained via the BCH formulas. 

The chain of biunivocal relations
\begin{equation}
\hat{\mathbf\Omega}\in\mathcal{G}/\mathcal{F} \Leftrightarrow 
\mathbf\Omega 
\in\mathcal{M} \Leftrightarrow \ket{\mathbf\Omega} \in \HH\,.
\end{equation}
is one of the most distinctive feature of the group-theoretic 
construction, as it establishes that
any GCS is univoquely associated to a point on $\mathcal{M}$, and 
viceversa. As a consequence, the invariant measure $d\mu(\hat{\mathbf \Omega})$ 
induces a measure $d\mu(\bm\Omega)$ upon ${\cal M}$. In fact,
it can be demonstrated \cite{ZhangFG90} that $\mathcal{M}$ is endowed 
with a natural metric that can be expressed in the $\eta_m$-coordinates 
as
\begin{equation}\label{natural metric}
ds^2=\sum_{mm'} g_{m m'}\,d\eta_m\, d{\eta_{m'}^*}\;\;\mbox{where} \;\;
 g_{m m'}\defi\frac
{\partial^2\log{\scalar{\tilde{\mathbf\Omega}}{\tilde{\mathbf \Omega}}}}
{\partial\eta_m\,\partial{\eta_{m'}^*}}~,
\end{equation}
with $\ket{\tilde{\mathbf \Omega}}\defi\ket{\mathbf \Omega}\!/N$ in \eqref{e.unnorm-GCS_inapp}.
After $ds^2$ one can define a canonical volume form on $\mathcal{M}$,  
i.e. the above mentioned measure on ${\cal M}$, via
\begin{equation}\label{natural measure}
d\mu(\mathbf \Omega)= 
\mbox{const} \times \det(g) \prod_{m} d\eta_m\, d{\eta_m^*}~.
\end{equation}
The manifold $\mathcal{M}$ 
is also equipped with a symplectic structure that allows one to identify 
it as a phase-space. In particular, the symplectic form on $\mathcal{M}$ 
has the coordinate representation
\begin{equation}\label{symplectic form}
\omega=-i\;\sum_{m m'}g_{m m'}\,d\eta_m\wedge d{\eta_{m'}^*}~,
\end{equation}
that can be used to define the Poisson brackets
\begin{equation} 
\left\{f,g\right\}_{PB}\defi i\;\sum_{m m'}g^{m m'}\left(\frac{\partial f}
{\partial \eta_m} \frac{\partial g}{\partial {\eta_{m'}^*}} - 
\frac{\partial f}{\partial \eta_{m'}^*} 
\frac{\partial g}{\partial {\eta_m}}\right)~,
\end{equation}
with $\sum_n g_{mn} g^{nm'}=\delta_m^{m'}$.

In the case of non-semisimple algebras, such as $\mathfrak{h}_4$ and 
$\mathfrak{h}_6$ for the harmonic and squeezed-harmonic oscillator, 
respectively, where a Cartan decomposition \eqref{e.GCS_Cartan_inapp} is not available, analogous decompositions 
exist, and the same procedure can be adopted.
This is explicitly done for $\mathfrak{h}_4$ at the end
of this material, where we show that the results are the same as those 
obtained in the semisimple case.

\subsection{PRECS: Parametric Representation with Environmental Coherent States} 

Parametric representations of composite systems can be built whenever a 
resolution of the identiy upon the Hilbert space of one of the 
subsystems is available. In Ref.~\cite{GiovannettiLM15}, for instance, 
the representation is introduced via $\int dx \ket{x}\bra{x}=\Id_\smc$, 
where $\ket{x}$ are the eigenstates of the position operator for one of 
two subsystems, and the integral is over the real axes. Our choice, 
which is pivotal to get to our final result, is based on the fact that 
parametric representations with GCS inherit from the group theoretic 
construction some properties that are essential in order to follow the 
quantum-to-classical crossover and formally define a classical limit of 
a quantum theory, according to the large-$N$ quantum 
approach.

The representation is defined as follows.
Consider an isolated bipartite system $\Psi=\clo+\Gamma$ with 
Hilbert space $\HH_\Psi=\HH_\smc\otimes\HH_\Gamma$, where $\Gamma$ is 
the principal system and $\clo$ its environment. The most general 
expression for a pure state of $\Psi$ is
\begin{equation}
\dket\Psi=\sum_{\gamma\xi}c_{\gamma\xi}\ket\gamma\otimes\ket\xi\quad
\mbox{with}\quad\sum_{\gamma\xi}|c_{\gamma\xi}|^2=1\,,
\label{e.dket_Psi_inapp}
\end{equation}
where $\lbrace\ket\gamma\rbrace_\Gamma$ and $\lbrace\ket\xi\rbrace_{\smc}$ 
are orthonormal bases for $\HH_\Gamma$ and $\HH_\smc$ respectively.
Inserting the above mentioned resolution of the identity upon 
$\HH_\smc$, for which we choose the one provided by GCS, Eq.~\eqref{e.id_resolution_inapp}, one gets
\begin{equation}\label{PRECS_inapp}
\dket{\Psi}= \int_{\mathcal{M}} {d\mu(\mathbf\Omega)\,\chi(\mathbf\Omega)\,
\ket{\mathbf\Omega}\otimes\ket{\phi(\mathbf\Omega)}}\,,
\end{equation}
where $\chi(\bm\Omega)$ is a function that 
can be chosen real, being defined via $\chi^2(\bm\Omega)\coloneqq\sum_\gamma
|\sum_\xi c_{\gamma\xi}\exval{\bm\Omega|\xi}|^2$. The element 
$\ket{\phi(\bm\Omega)}$ of $\HH_\Gamma$ is normalized, and hence 
describe a pure state of $\Gamma$. Due to the normalization of 
$\dket{\Psi}$, it is
\begin{equation}\label{e.chi2}
\int_{\cal M} d\mu(\mathbf\Omega)\chi^2(\mathbf\Omega)=1\,,
\end{equation}
meaning that $\chi^2(\Omega)$ can be interpreted as a probability 
distribution on $\cal M$. 
The above expressions have a clear physical 
interpretation:
reminding that each point ${\bm\Omega}\in\cal M$ is in 
one-to-one correspondence with a GCS $\ket{\bm\Omega}\in\HH_\smc$,
we can say that $\ket{\phi(\bm\Omega)}$ is the state of $\Gamma$ 
conditioned to $C$ being in the GCS $\ket{\bm\Omega}$, a circumstance 
that occurs with probability $\chi^2(\bm\Omega)$ when $\Psi$ is in the pure state $\dket{\Psi}$.
This interpretation is consistent with the following 
relations~\cite{CalvaniPhD13}
\begin{equation}\label{e.PRECS_chi2}
\chi^2(\mathbf{\Omega})=
\bra{\mathbf{\Omega}}\rho_\smc\ket{\mathbf{\Omega}}~,
\end{equation}
and
\begin{equation}\label{e.PRECS_rho}
\rho_{\Gamma}=
\int_{\mathcal{M}} d\mu(\mathbf{\Omega}) 
\chi(\mathbf{\Omega})^{2}\proj{\phi(\mathbf{\Omega})}~,
\end{equation}
where $\rho_{\Gamma(C)}\coloneqq{\rm Tr}_{C(\Gamma)}\dket{\Psi}\!\dbra{\Psi}$.
Notice that the diagonal-like form \eqref{e.PRECS_rho} of $\rho_\Gamma$ 
is not generally granted for parametric representations such that the 
identity resolution is in terms of non-orthogonal states, as in the GCS 
case. In fact, it is the specific overcompletenes of GCS
that ensures Eq.~\eqref{e.PRECS_rho} to hold.

Finally, it is important to remind that despite parametric 
representations allow
one to use pure states $\ket{\phi(\bm\Omega)}$ to describe $\Gamma$, 
this should by no means be intended as if $\Gamma$ were in a pure state.
In fact, due to the parametric dependence of $\ket{\phi(\bm\Omega)}$ on 
$\bm\Omega$, the density operator $\rho_\Gamma$ in 
Eq.~\eqref{e.PRECS_rho} is not a projector, reflecting that 
$C$ and $\Gamma$ are entangled, as far as the form 
\eqref{e.dket_Psi_inapp} of $\dket{\Psi}$ stays general.
To this respect, it is easily verified that when $\dket{\Psi}$ is 
separable the above parametric dependence dies out.

\subsection{Derivation of the symbol of $\hat H_\smc$}

In this part we express the symbol $\bra{\lambda}\hat{H}_\smc\ket{\lambda}$ of the clock-Hamiltonian, 
as introduced in Sec.~\ref{s.letscallit-t} of the paper, in terms of
the complex parameter $\lambda:=\Omega_\ell=\varrho e^{-i\varphi}$ 
that defines the GCS $\ket{\lambda}$ via 
$\ket{\lambda}\defi e^{\hat{W}}\ket{G}$ with
$\ket{G}$ the clock reference state satisfying $\hat{R}_m\ket{G}=0~\forall m$, 
$\hat{D}_\delta\ket{G}=g_\delta\ket{G}$, and
$\hat W:=\hat{W}_\ell:=\Omega_\ell
\hat{R}_\ell^\dagger-\Omega_\ell^* \hat{R}_\ell$.
Recalling that
$\hat{H}_{\smc}=\varsigma\hat{D}_1+K$ with $K\defi-\varsigma g_1$ and $\varsigma^2=\pm 1$ such that $\epsilon\defi\varsigma d_{1\ell}$ is real
and positive, we write
\begin{align}
&\bra{\lambda}\hat{H}_\smc\ket{\lambda}=
K+\varsigma\bra{G}e^{-\hat{W}}\hat{D}_1 e^{\hat{W}}\ket{G}\nonumber\\
=&\;K+\varsigma\bra{G}\hat{D}_1+[\hat{W},\hat{D}_1]+\frac{1}{2!}[\hat{W},[\hat{W},\hat{D}_1]]+\nonumber\\
&+\frac{1}{3!}[\hat{W},[\hat{W},[\hat{W},\hat{D}_1]]]+ \nonumber\\
&+\frac{1}{4!}[\hat{W},[\hat{W},[\hat{W},[\hat{W},\hat{D}_1]]]]+...\ket{G}\nonumber\\
=&\;K+\varsigma\bra{G}\hat{D}_1+[\hat{W},\hat{D}_1]+\sum_{\delta}\bigg[\frac{1}{2!}(-2d_{1 l}d_{\delta \ell}\varrho^2 \hat{D}_\delta)+\nonumber\\
&+\frac{1}{3!}(-2d_{1 \ell}d_{\delta \ell}\varrho^2[\hat{W},\hat{D}_\delta])+ \nonumber \\
&+\frac{1}{4!}\sum_{\theta}(-2d_{1 \ell}d_{\theta \ell}\varrho^2)(-2d_{\theta \ell}d_{\delta \ell}\varrho^2 \hat{D}_\delta)+...\bigg]\ket{G}\nonumber\\
=&\;\varsigma\sum_{\delta} g_\delta\bigg[\frac{1}{2!}(-2\varrho^2d_{1 \ell} d_{\delta \ell})+\nonumber\\
&+\frac{1}{4!}\sum_{\theta}(-2\varrho^2d_{1 \ell}d_{\theta \ell})(-2\varrho^2d_{\theta \ell} 
d_{\delta \ell})+...\bigg]\nonumber\\
=&\;\varsigma d_{1\ell}\sum_{\delta} g_\delta d_{\delta \ell}
\bigg(\sum_{n=1}^{\infty}\frac{(-1)^n}{(2n)!}(\sqrt{2}
\varrho)^{2n}a^{2n-2}\bigg)\nonumber\\
=&\;\epsilon a^{-2}\varsigma^2 b^2
\left(\cos\Big(a\sqrt{2}\varrho\Big)-1\right):=H_\smc(\varrho)~,
\end{align}
where $a^2=\sum_\theta d_{\theta\ell}^2$ and 
$\varsigma^2 b^2=\sum_\delta g_\delta d_{\delta\ell} $. 
For the sake of a lighter notation, in what follows and in the main work 
we set $\varsigma^2 a^2=2$, which means that the raising
and diagonal operators are multiplied by
$\sqrt{2\varsigma^2}/a$, and their eigenvectors are rescaled
accordingly. We thus finally get
\begin{align}
&\bra{\lambda}\hat{H}_\smc\ket{\lambda}=\frac{\epsilon b^2}{2}
(\cos(2\varsigma\varrho)-1)~.
\end{align}

\subsection{The pullback-by-$F$ and the Poisson brackets on $\mathcal{M}_\smc$}

In this part we will explicitly calculate the Poisson brackets 
$\{\cdot,\cdot\}_\smc$ induced on $\mathcal{M}_\smc$ via the 
pullback-by-$F$. We recall that, given the manifolds
$\mathcal{M}_\smc$ and $\mathcal{M}_\Gamma$ for the clock $\clo$ and the 
evolving system $\Gamma$ as from the GCS construction, the map 
$F:U_\smc\subset\mathcal{M}_\smc\rightarrow U_\Gamma \subset 
\mathcal{M}_\Gamma$, is defined as
\begin{equation}\label{map F}
\begin{dcases}
q_j=\sqrt{2}\, b\, \varsigma\,\sin\left(\varsigma\varrho\right)\cos(\varphi)\,v_j~,\\
p_j=\frac{\sqrt{2} \,b}{\varsigma}\,\sin\left(\varsigma\varrho\right)\sin(\varphi)\,v_j~,
\end{dcases}
\end{equation}
with $\sum_j {v_j^2}=1$. We remind that
the Poisson brackets are defined on a generic symplectic 
manifold $\mathcal{M}$ starting from its symplectic form 
$\omega=\frac{1}{2}\sum_{\mu\nu}\omega_{\mu\nu}\,dx^\mu\wedge dx^\nu$, 
via $\{f,g\}=\sum_{\mu\nu}\omega^{\mu\nu}\,\partial_{x_\mu} 
f\,\partial_{x_\nu} g$ with $\sum_{\sigma} 
\omega_{\mu\sigma}\,\omega^{\sigma\nu}=\delta_\mu^\nu$, and 
$x^\mu$ $\left(\mu=1,...,2n=\mathrm{dim}\mathcal{M}\right)$, 
$f,\,g$ are some coordinates and generic functions on $\mathcal{M}$, respectively. 
In fact, the Darboux theorem guarantees that there exists local coordinates $x^{\mu}=(q_1,...,q_n,p_1,...,p_n)$ 
such that $\omega=\mathfrak{h}\,\sum_{j=1}^n {dp_j \wedge dq_j}$ and 
$\{f,g\}=\mathfrak{h}^{-1}\;\sum_{j=1}^n \left(\partial_{q_j} 
f\;\partial_{p_j} g-\partial_{p_j} f\;\partial_{q_j} g\right)$, with 
$\mathfrak{h}=const.$.
This said, being
$(q_j,p_j)$ in Eq.~\eqref{map F} Darboux coordinates, i.e. 
$\{q_j,p_j\}_\Gamma=\mathfrak{h}^{-1}\delta_{ij}$, the symplectic 
form $\omega_\Gamma$ on $U_\Gamma \subset \mathcal{M}_\Gamma$ is 
\begin{equation} \omega_\Gamma=\mathfrak{h}\,\sum_j {dp_j \wedge dq_j}~.
\end{equation} 
We can now calculate the pullback-by-$F$ of 
$\omega_\Gamma$ as 
\begin{align} 
(\omega_\Gamma)^*&=&\mathfrak{h}\,\sum_j &\left[\sqrt{2} b \, 
\cos\left(\varsigma\varrho\right)\sin(\varphi)\,v_j\;d\varrho\right. 
\nonumber\\
                &&                            &+\left.\frac{\sqrt{2}b}{\varsigma}\,
                \sin\left(\varsigma\varrho\right)\cos(\varphi)\,v_j\;d\varphi\right] \nonumber\\
                &&                       \wedge&\left[\sqrt{2}\,b \varsigma^2  \, 
               \cos\left(\varsigma\varrho\right)\cos(\varphi)\,v_j\;d\varrho\right. \nonumber\\
                        &&														 &-\left.\sqrt{2} b \varsigma\,
							\sin\left(\varsigma\varrho\right)\sin(\varphi)\,v_j\;d\varphi\right] \nonumber\\
				&=&\mathfrak{h}\,\sum_j &\left[-b^2\, \varsigma\,\sin\left(2\varsigma\varrho\right)\,v_j^2\right. \sin^2{\varphi}\;\, d\varrho\wedge d\varphi  \nonumber\\
							&&                       &+b^2\, \varsigma\,\sin\left(2\varsigma\varrho\right)\,v_j^2  \left.\;\;\;\cos^2{\varphi}\;\, d\varphi \wedge d\varrho\right]\nonumber\\
							&=&&\!\!\!\!\!\!\!\!\!\!\!\!\!\!\mathfrak{h} b^2 \varsigma\,\sin\left(2\varsigma \varrho\right)\;\,d\varphi \wedge d\varrho~. 
\label{pullback of omega}		     
\end{align}     
Finally $(\omega_\Gamma)^*$ defines Poisson brackets on 
$\mathcal{M}_\smc$ via
\begin{align}
\{f_\smc,g_\smc\}_\smc=\;&\frac{1}{\mathfrak{h}b^2\,\varsigma \,
\sin\left(2\varsigma \varrho\right)}\nonumber\\
                        &\left(\frac{\partial f_\smc}{\partial\varrho}\frac{\partial g_\smc}{\partial \varphi}-\frac{\partial f_\smc}{\partial \varphi}\frac{\partial g_\smc}{\partial\varrho}\right)~.
\end{align}
We clarify that our choice \eqref{map F} for the map $F$
follows from the one suggested in Ref.~\cite{ZhangFG90}, but other 
choices are possible.
% and have already been used by some of the authors, leading to results that are fully consistent with those reported here.

\subsection{The Heisenberg algebra $\mathfrak{h}_4$}

When the Lie algebra $\mathfrak{g}$, to which the clock Hamiltonian 
$\hat{H}_\smc$ belongs, is semisimple, the GCS are built 
starting from the Cartan decomposition. However a similar construction 
can be put forward for the non-semisimple algebra $\mathfrak{h}_4$. The 
latter is defined by the set $\{\hat{n}=\hat{a}^\dagger \hat{a}, 
\hat{a}, \hat{a}^\dagger, \hat{\mathbb{I}}\}$ with commutation 
relations 
$[\hat{a},\hat{a}^\dagger]=\hat{\mathbb{I}},\;\;[\hat{a},\hat{\mathbb{I}}]=[\hat{a}^\dagger,\hat{\mathbb{I}}]=0$. 
The GCS $\ket{\alpha}$, usually called harmonic-oscillator coherent 
states or just coherent states, are in one-to-one correspondence with 
the points of the complex plane $\mathbb{C}$ and can be equivalently 
defined as $\ket{\alpha}=e^{\alpha 
\hat{a}^\dagger-\alpha^*\hat{a}}\Ket{G}=e^{-|\alpha|^2/2}\,e^{\alpha 
\hat{a}^\dagger}\Ket{G}$ with $\hat{n}\Ket{G}=\hat{a}\Ket{G}=0$, or as 
$\hat{a}\ket{\alpha}=\alpha\ket{\alpha}$ with $\alpha\in\mathbb{C}$. 
When the clock $\clo$ admits a proper classical limit, the 
Schr\"{o}dinger equation for the evolving system $\Gamma$ can be 
obtained, as shown in the main work, implementing the PaW mechanism via the 
PRECS and considering a fixed GCS $\ket{\lambda}=e^{\lambda 
\hat{a}^\dagger-\lambda^*\hat{a}}\Ket{G}=N_\varrho e^{\lambda 
\hat{a}^\dagger}\Ket{G}$ with $\lambda=\varrho e^{i\varphi}$, for which, 
being $[\hat{n}, 
e^{\lambda^*\hat{a}}]=-\lambda^*\hat{a}e^{\lambda^*\hat{a}}$, it is 
$\bra{\lambda}\hat{H}_\smc\ket{\alpha}=i\epsilon\frac{d}{d 
\varphi}\scalar{\lambda}{\alpha}$, where 
$\hat{H}_\smc=\epsilon \hat{n}$. Again, the temporal parameter $t^{\rm QM}$ 
turns out to be $t^{\rm QM}=(\hbar/\epsilon) \varphi$. Moreover, since it is 
trivial to show that 
$H_\smc(\lambda)=\Bra{\lambda}\hat{H}_\smc\Ket{\lambda}=\epsilon 
\varrho^2$, the considerations concerning the parameter $\varrho$ and 
the stationary Schr\"{o}dinger equation for $\Gamma$ still apply. 
For what concerns the uncertainty relation, a phase-operator can be 
defined via $\hat{a}=\hat{n}^{1/2}e^{i\hat{\phi}}$. Finally, when 
$\Gamma$ becomes macroscopic and presents a completely classical 
behaviour, its dynamics is ruled by the Hamilton equations according to 
a temporal parameter $t^{\rm CL}=t^{\rm QM}$. This result can be obtained 
following the same line of reasoning of the main work and choosing the 
map $F$ to be $q_j-ip_j=v_j\sqrt{2}\varrho e^{i\varphi}$ with $\sum_j 
v_j^2=1$.

\end{document}